\documentclass[9pt,twocolumn,twoside, lineno]{pnas-blank}
\templatetype{pnasresearcharticle} 
\usepackage{mathpazo}
\usepackage{xcolor}
\usepackage[normalem]{ulem}

\usepackage{float}
\usepackage{xparse}
\usepackage{amsmath}
\newcommand{\refst}{^\mathrm{(wt)}}
\newcommand{\perst}{^\mathrm{(mut)}}
\newcommand{\pact}{p_\text{act}}

\title{The Energetics of Molecular Adaptation in Transcriptional Regulation}
\author[a]{Griffin Chure}
\author[a]{Manuel Razo-Mejia}
\author[a,b]{Nathan M. Belliveau}
\author[c]{Tal Einav}
\author[a]{Zofii A. Kaczmarek}
\author[a,d]{Stephanie L. Barnes}
\author[e]{Mitchell Lewis}
\author[a, f, 1]{Rob Phillips} 

\affil[a]{Department of Biology and Biological Engineering, California Institute of Technology, Pasadena, CA, USA}
\affil[b]{Present address: Howard Hughes Medical Institute and Department of Biology, University of Washington, Seattle, WA, USA}
\affil[c]{Department of Physics, California Institute of Technology, Pasadena, CA, USA}
\affil[d]{Present Address: Department of Science and Technology, Windward School, Los Angeles, CA, USA}
\affil[e]{Department of Biochemistry and Molecular Biophysics, University of Pennsylvania School of Medicine, Philadelphia, PA, USA}
\affil[f]{Division of Physics, Mathematics and Astronomy, California Insitute of Technology, Pasasdena, CA, USA}

\leadauthor{Chure} 

\authordeclaration{The authors declare no conflict of interest.}
\correspondingauthor{\textsuperscript{1}To whom correspondence should be addressed. E-mail: phillips@pboc.caltech.edu}
\keywords{Transcriptional Regulation $|$ Allostery $|$ MWC Model $|$ Statistical Mechanics $|$ Epistasis $|$ Biophysics $|$ Evolution $|$ Mutation} 

\begin{abstract}
Mutation is a critical mechanism by which evolution explores the functional
landscape of proteins. Despite our ability to experimentally inflict
mutations at will, it remains difficult to link sequence-level perturbations
to systems-level responses. Here, we present a framework centered on
measuring changes in the free energy of the system to link individual
mutations in an allosteric transcriptional repressor to the parameters which
govern its response. We find the energetic effects of the mutations can be
categorized into several classes which have characteristic curves as a
function of the inducer concentration. We experimentally test these
diagnostic predictions using the well-characterized LacI repressor of
\textit{Escherichia coli}, probing several mutations in the DNA binding and
inducer binding domains. We find that the change in gene expression due to a
point mutation can be captured by modifying only a subset of the model
parameters that describe the respective domain of the wild-type protein.
These parameters appear to be insulated, with mutations in the DNA binding
domain altering only the DNA affinity and those in the inducer binding domain
altering only the allosteric parameters. Changing these subsets of parameters
tunes the free energy of the system in a way that is concordant with
theoretical expectations. Finally, we show that the induction profiles and
resulting free energies associated with pairwise double mutants can be predicted with
quantitative accuracy given knowledge of the single mutants, providing an
avenue for identifying and quantifying epistatic interactions.
\end{abstract}

\significancestatement{ 
We present a biophysical model of allosteric transcriptional regulation that
directly links the location of a mutation within a repressor to the
biophysical parameters that describe its behavior. We explore the phenotypic
space of a repressor with mutations in either the inducer binding or DNA binding
domains. Using the LacI repressor in \textit{E. coli}, we make sharp, falsifiable
predictions and use this framework to generate a null hypothesis for how
double mutants behave given knowledge of the single mutants. Linking
mutations to the parameters which govern the system allows for quantitative
predictions of how the free energy of the system changes as a result,
permitting coarse graining of high-dimensional data into a single-parameter
description of the mutational consequences.
\include{significance}
}

\begin{document}
\maketitle
\thispagestyle{firststyle}
\ifthenelse{\boolean{shortarticle}}{\ifthenelse{\boolean{singlecolumn}}{\abscontentformatted}{\abscontent}}{}

\dropcap{T}hermodynamic treatments of transcriptional regulation have been
fruitful in their ability to generate quantitative predictions of
gene expression as a function of a minimal set of physically meaningful
variables \cite{Ackers1982, Buchler2003, Vilar2003, Garcia2011,
Daber2011a,Brewster2014, Weinert2014, Rydenfelt2014, Razo-Mejia2014,
Razo-Mejia2018, Bintu2005, Bintu2005a, Kuhlman2007}. These models
quantitatively describe numerous properties of input-output functions, such
as the leakiness, saturation, dynamic range, steepness of response, and
the [$EC_{50}$] -- the concentration of inducer at which the response is half
maximal. The mathematical forms of these phenotypic properties are couched in
terms of a minimal set of experimentally accessible variables, such as the
inducer concentration, transcription factor copy number, and the DNA sequence
of the binding site \cite{Razo-Mejia2018}. While the amino acid sequence of
the transcription factor is another controllable variable, it is
seldom implemented in quantitative terms considering mutations with subtle changes
in chemistry frequently result in unpredictable physiological
consequences. In this work, we examine how a series of
mutations in either the DNA binding or inducer binding domains of a
transcriptional repressor influence the values of the biophysical parameters
which govern its regulatory behavior.

We first present a theoretical framework for understanding how mutations in
the repressor affect different parameters and alter the free energy of the
system. The multi-dimensional parameter space of the aforementioned
thermodynamic models is highly degenerate with multiple combinations of
parameter values yielding the same phenotypic response. This degeneracy can
be subsumed into the free energy of the system, transforming the input-output
function into a one-dimensional description with the form of a Fermi function \cite{Swem2008,Keymer2006}.
 We find that the parameters capturing the allosteric nature of
the repressor, the repressor copy number, and the DNA binding specificity contribute
independently to the free energy of the system with different degrees of
sensitivity. Furthermore, changes restricted to one of these
three groups of parameters result in characteristic changes in the free
energy relative to the wild-type repressor, providing falsifiable
predictions of how different classes of mutations should behave.

Next, we test these descriptions experimentally using the well-characterized
transcriptional repressor of the \textit{lac} operon LacI in \textit{E.
coli} regulating expression of a fluorescent reporter. We introduce a series
of point mutations in either the inducer binding or DNA binding domain. We
then measure the full induction profile of each mutant, determine the minimal
set of parameters that are affected by the mutation, and predict
how each mutation tunes the free energy at different inducer concentrations,
repressor copy numbers, and DNA binding strengths.
We find in general that mutations in the DNA binding domain only influence
DNA binding strength, and that mutations within the inducer binding domain
affect only the parameters which dictate the allosteric response. The
degree to which these parameters are insulated is notable, as the very nature
of allostery suggests that all parameters are intimately connected, thus enabling
binding events at one domain to be "sensed" by another. 

With knowledge of how a collection of DNA binding and inducer binding single mutants
behave, we predict the induction profiles and the free energy
changes of pairwise double mutants with quantitative accuracy. We find that
the energetic effects of each individual mutation are additive, indicating that
epistatic interactions are absent between the mutations examined here. Our model
provides a means for identifying and quantifying the extent of epistatic
interactions in a more complex set of mutations, and can shed light on how the
protein sequence and general regulatory architecture coevolve.

 \section*{Results}
\begin{figure*}[h]
     \centering
     \includegraphics{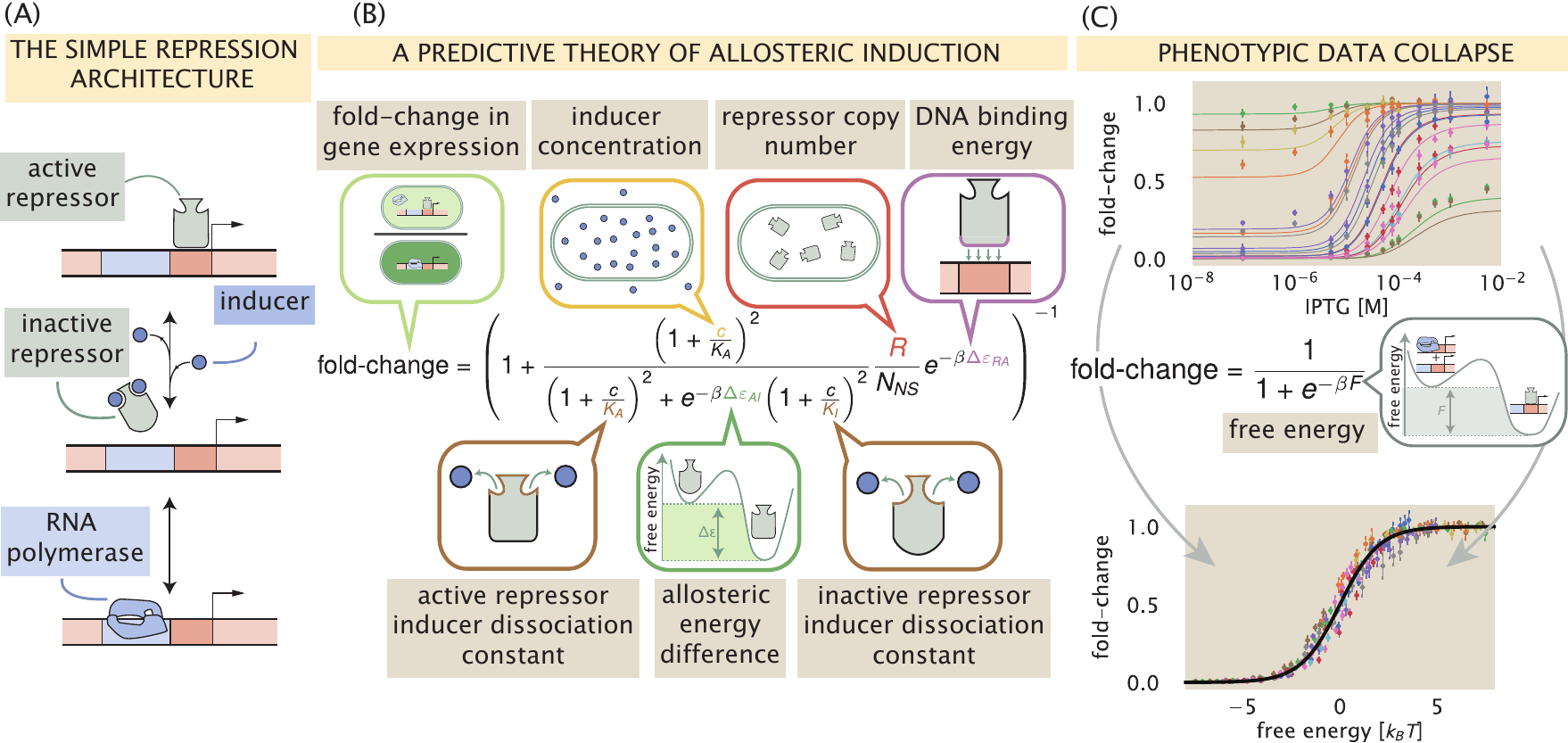}
     \caption{A predictive framework for phenotypic and energetic dissection of
 the simple repression motif. (A) The inducible simple repression
 architecture. When in the active state, the repressor (gray) binds the
 cognate operator sequence of the DNA (red box) with high specificity,
 preventing transcription by occluding binding of the RNA polymerase to the
 promoter (blue rectangle). Upon addition of an inducer molecule, the inactive
 state becomes energetically preferable and the repressor no longer binds the operator
 sequence with appreciable specificity. Once unbound from the operator,
 binding of the RNA polymerase (blue) is no longer blocked and transcription
 can occur. (B) The simple repression input-output function for an allosteric
 repressor with two inducer binding sites. The key parameters are identified in
 speech bubbles. (C) Fold-change in gene expression collapses as a function of the
 free energy. The input-output function in (B) can be re-written as
 a Fermi function with an energetic parameter $F$ which is the energetic
 difference between the repressor bound and unbound states of the promoter.
 Top panel shows induction profiles reported in Razo-Mejia \textit{et al.} 2018
 \cite{Razo-Mejia2018} of eighteen different strains over twelve concentrations
 of the gratuitous inducer Isopropyl $\beta$-D-1-thiogalactopyranoside (IPTG). Upon calculation of the
 free energy, the data collapse onto a single master curve (bottom panel)
 defined by $F$.}
 \label{fig:induction_theory}
 \end{figure*}

 This work considers the inducible simple repression regulatory motif
 [depicted in Fig. \ref{fig:induction_theory}(A)] from a thermodynamic
 perspective which has been thoroughly dissected and tested experimentally
 \cite{Garcia2011, Brewster2014, Razo-Mejia2018}. While we direct the reader
 to the SI text for a complete derivation, the result of this extensive
 theory-experiment dialogue is a succinct input-output function [schematized
 in Fig. \ref{fig:induction_theory}(B)] that computes the fold-change in gene
 expression relative to an unregulated promoter. This function is of the form
 \begin{equation}
 \text{fold-change} = \left(1 + {R_A \over
 N_{NS}}e^{-\beta\Delta\varepsilon_{RA}}\right)^{-1},
 \label{eq:foldchange}
 \end{equation}
 where $R_A$ is the number of active repressors per cell, $N_{NS}$ is the
 number of non-specific binding sites for the repressor,
 $\Delta\varepsilon_{RA}$ is the binding energy of the repressor
 to its specific binding site relative to the non-specific background, and
 $\beta$ is defined as ${1 \over k_B T}$ where $k_B$ is the Boltzmann constant
 and $T$ is the temperature. While this theory requires knowledge of the number
 of \textit{active} repressors, we often only know the total number $R$
 which is the sum total of active and inactive repressors. We can define a
 prefactor $p_\text{act}(c)$ which captures the
 allosteric nature of the repressor and encodes the probability a
 repressor is in the active (repressive) state rather than the inactive state
 for a given inducer concentration $c$, namely,
 \begin{equation}
 p_\text{act}(c) = {\left(1 + {c \over K_A}\right)^n \over \left(1 + {c \over
 K_A}\right)^n + e^{-\beta\Delta\varepsilon_{AI}}\left(1 + {c \over
 K_I}\right)^n}.
 \label{eq:pact}
 \end{equation}
 Here, $K_A$ and $K_I$ are the dissociation constants of the inducer to the
 active and inactive repressor, $\Delta\varepsilon_{AI}$ is the energetic
 difference between the repressor active and inactive states, and $n$ is the number of
 allosteric binding sites per repressor molecule ($n=2$ for LacI). With this in
 hand, we can define $R_A$ in \eqref{eq:foldchange} as $R_A = p_\text{act}(c)
 R$.

 A key feature of \eqref{eq:foldchange} and \eqref{eq:pact} is that the diverse
 phenomenology of the gene expression induction profile can be collapsed onto
 a single master curve by rewriting the input-output function in terms of the
 free energy $F$ [also called the Bohr parameter \cite{Phillips2015}],
 \begin{equation}
 \text{fold-change} = \left(1 + e^{-\beta F}\right)^{-1},
 \label{eq:collapse}
 \end{equation}
 where
 \begin{equation}
 F = -k_BT \log\pact(c) - k_BT\log\left({R \over N_{NS}}\right) + \Delta\varepsilon_{RA}.
 \label{eq:Bohr}
 \end{equation}
 Hence, if different combinations of parameters yield the same free energy,
 they will give rise to the same fold-change in gene expression, enabling us
 to collapse multiple regulatory scenarios onto a single curve. This can be
 seen in Fig.~\ref{fig:induction_theory}(C) where eighteen unique inducer
 titration profiles of a LacI simple repression architecture collected and
 analyzed in Razo-Mejia \textit{et al.} 2018 \cite{Razo-Mejia2018} collapse onto a
 single master curve. The tight distribution about this curve reveals that
 fold-change across a variety of genetically distinct individuals can be
 adequately described by a small number of parameters. Beyond predicting the
 induction profiles of different strains, the method of data collapse inspired
 by \eqref{eq:collapse} and \eqref{eq:Bohr} can be used as a tool to identify
 mechanistic changes in the regulatory architecture \cite{Swem2008}. Similar data collapse approaches
 have been used previously in such a manner and have proved vital for
 distinguishing between changes in parameter values and changes in the
 fundamental behavior of the system \cite{Swem2008, Keymer2006}.
 
 Assuming that a given mutation does not result in a non-functional protein, it
 is reasonable to say that any or all of the parameters in \eqref{eq:foldchange}
 can be affected by the mutation, changing the observed induction profile and therefore the free 
 energy. To examine how the free energy of a mutant $F\perst$
 differs from that of the wild-type $F\refst$, we define $\Delta F = F\perst -
 F\refst$, which has the form
 \begin{equation}
     \begin{aligned}
 \Delta F = -k_BT\log\left({p_\text{act}\perst(c) \over  p_\text{act}\refst(c)}\right) &- k_BT \log\left({R\perst \over R\refst}\right)\\
  &+ (\Delta\varepsilon_{RA}\perst - \Delta\varepsilon_{RA}\refst).
     \end{aligned}
     \label{eq:delF}
 \end{equation}
 
\begin{figure*}[t]
         \centering
         \includegraphics{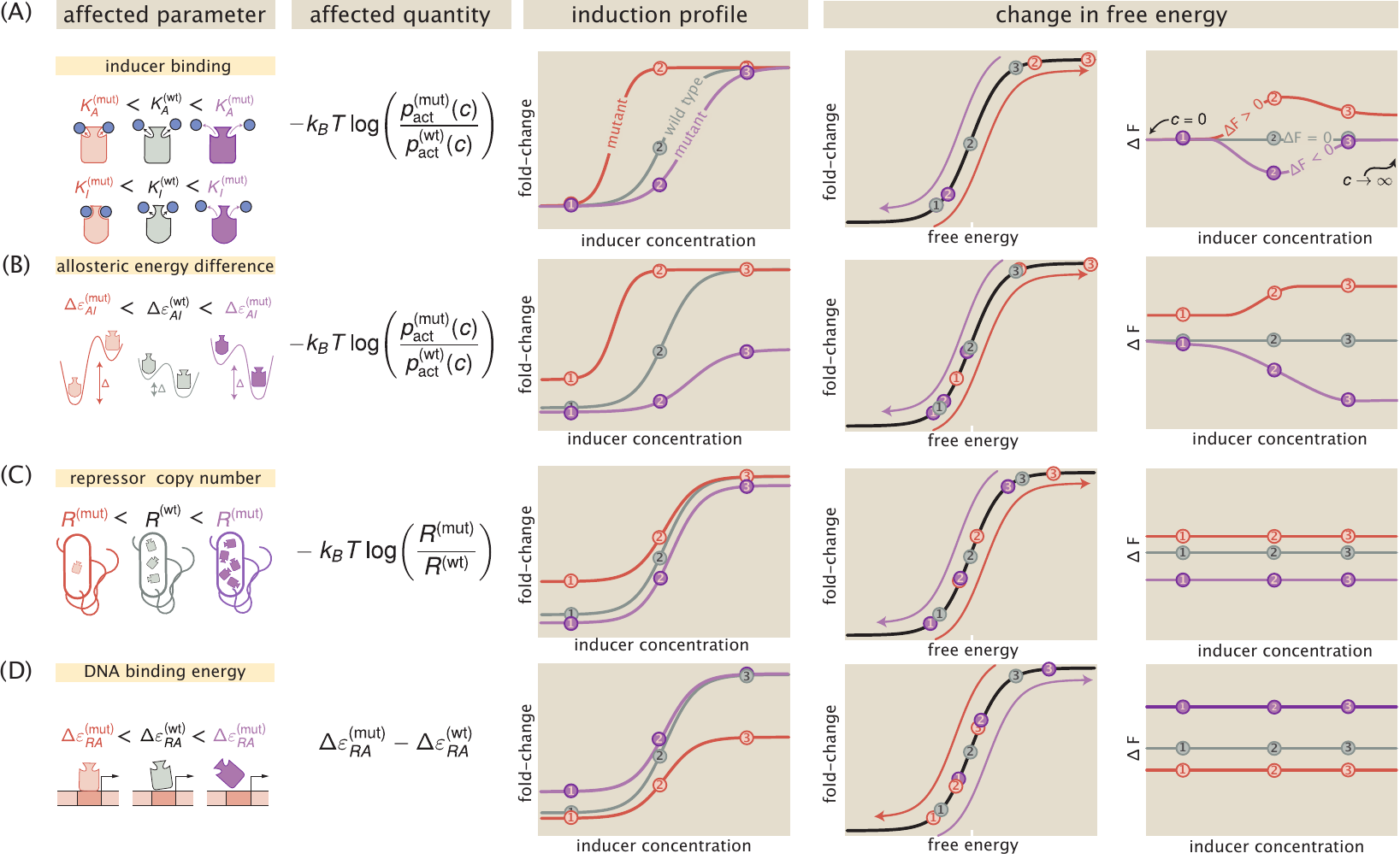}
     \caption{Parametric changes due to mutations alter the free energy.  The first column schematizes
     the changed parameters and the second column reflects which quantity in
     \eqref{eq:delF} is affected. The third column shows representative induction
     profiles from mutants which have smaller (red) and larger (purple) values for
     the parameters than the wild-type (grey). The fourth and fifth columns illustrate how the
     free energy is changed as a result. Purple and red arrows indicate the direction
     in which the points are translated about the master curve. Three concentrations
     (points labeled 1, 2, and 3) are shown  to illustrate how each point is
     moved in free energy space.}
     \label{fig:deltaF_theory}
 \end{figure*}

 $\Delta F$ describes how a mutation translates a point across the master
 curve shown in Fig. \ref{fig:induction_theory}(C). As we will show in the
 coming paragraphs [illustrated in Fig.
 \ref{fig:deltaF_theory}], this formulation coarse grains the myriad
 parameters shown in \eqref{eq:foldchange} and \eqref{eq:pact} into three
 distinct quantities, each with different sensitivities to parametric changes. By
 examining how a mutation changes the free energy changes as a function of the inducer
 concentration, one can draw conclusions as to which parameters have been
 modified based solely on the shape of the curve.
 To help the reader understand how various perturbations to the parameters
 tune the free energy, we have hosted an interactive figure on the paper website
 \href{http://www.rpgroup.caltech.edu/mwc_mutants} which makes
 exploration of parameter space a simpler task.

 The first term in \eqref{eq:delF} is the log ratio of the probability of a
 mutant repressor being active relative to the wild type at a given inducer concentration $c$. This quantity defines how
 changes to any of the allosteric parameters -- such as inducer binding
 constants $K_A$ and $K_I$, or active/inactive state energetic difference
 $\Delta\varepsilon_{AI}$ -- alter the free energy $F$, which can be interpreted
 as the free energy difference between the
 repressor bound and unbound states of the promoter. Fig.
 \ref{fig:deltaF_theory} (A) illustrates how changes to the inducer binding
 constants $K_A$ and $K_I$ alone alter the induction profiles and resulting
 free energy as a function of the inducer concentration. In the limit where $c
 = 0$, the values of $K_A$ and $K_I$ do not factor into the calculation of
 $p_\text{act}(c)$ given by \eqref{eq:pact}, meaning that
 $\Delta\varepsilon_{AI}$ is the lone parameter setting the residual activity
 of the repressor. Thus, if only $K_A$ and $K_I$ are altered by a mutation,
 then $\Delta F$ should be $0\, k_BT$ when $c = 0$, illustrated by the overlapping
 red, purple, and grey curves in the right-hand plot of Fig.
 \ref{fig:deltaF_theory}(A). However, if $\Delta\varepsilon_{AI}$ is
 influenced by the mutation (either alone or in conjunction with $K_A$ and
 $K_I$), the leakiness will change, resulting in a non-zero $\Delta F$ when
 $c=0$. This is illustrated in Fig. \ref{fig:deltaF_theory} (B) where
 $\Delta\varepsilon_{AI}$ is the only parameter affected by the
 mutation. 
 
 It is important to note that for a mutation which perturbs only the inducer binding constants, the
 dependence of $\Delta F$ on the inducer concentration can be non-monotonic. While the precise values of $K_A$ and $K_I$
 control the sensitivity of the repressor to inducer concentration, it
 is the ratio $K_A / K_I$ that defines whether this non-monotonic behavior is
 observed. This can be seen more clearly when we consider the limit of saturating
 inducer concentration,
 \begin{equation}
     \lim\limits_{c \rightarrow \infty} \log\left({p_\text{act}\perst \over p_\text{act}\refst}\right) \approx \log\left[{1 + e^{-\beta\Delta\varepsilon_{AI}\refst} \left({K_A\refst \over K_I\refst}\right)^n \over 1 + e^{-\beta\Delta\varepsilon_{AI}\refst} \left({K_A\perst \over K_I\perst}\right)^n}\right]
     \label{eq:kaki_sat_c},
 \end{equation}
 which illustrates that $\Delta F$ returns to zero at saturating inducer concentration when $K_A / K_I$ is the same for both the
 mutant and wild-type repressors, so long as $\Delta\varepsilon_{AI}$ is
 unperturbed. Non-monotonicity can \textit{only} be achieved by changing $K_A$
 and $K_I$ and
 therefore serves as a diagnostic for classifying mutational effects reliant
 solely on  measuring the change in free energy.
 
 The second term in \eqref{eq:delF} captures how changes in the repressor
 copy number contributes to changes in free energy. It is important to note that this
 contribution to the free energy change depends on the total number of
 repressors in the cell, not just those in the active state. This emphasizes
 that changes in the expression of the repressor are energetically divorced
 from changes to the allosteric nature of the repressor. As a consequence, the
 change in free energy is constant for all inducer concentrations, as is
 schematized in Fig. \ref{fig:deltaF_theory}(C). Because magnitude of the change
 in free energy scales logarithmically with changing repressor copy number, a mutation
 which increases expression from 1 to 10 repressors per cell is more impactful
 from an energetic standpoint ($k_BT \log(10) \approx 2.3\,  k_BT$) than an
 increase from 90 to 100 ($k_BT \log(100/90) \approx 0.1\, k_BT$). Appreciable
 changes in the free energy only arise when variations in the repressor copy
 number are larger than or comparable to an order of magnitude. Changes of
 this magnitude are certainly possible from a single point mutation, as it has
 been shown that even synonymous substitutions can drastically change
 translation efficiency \cite{Frumkin2018}.
 
 The third and final term in \eqref{eq:delF} is the difference
 in the DNA binding energy between the mutant and wild-type repressors. All else
 being equal, if the
 mutated state binds more tightly to the DNA than the wild type
 ($\Delta\varepsilon_{RA}\refst > \Delta\varepsilon_{RA}\perst$), the net change
 in the free energy is negative, indicating that the repressor bound states
 become more energetically favorable due to the mutation.
 Much like in the case of changing repressor copy number, this quantity
 is independent of inducer concentration and is therefore also constant
 [Fig. \ref{fig:deltaF_theory}(D)]. However, the
 magnitude of the change in free energy is linear with DNA binding affinity 
 while it is logarithmic with respect to changes in the repressor copy number. Thus, to
 change the free energy by $1\, k_BT$, the repressor copy number must change
 by a factor of $\approx 2.3$ whereas the DNA binding energy must change by $1\, k_BT$. 
 
 The unique behavior of each quantity in \eqref{eq:delF} and its sensitivity with
 respect to the parameters makes $\Delta F$ useful as a diagnostic tool to classify
 mutations. Given a set of fold-change measurements, a simple rearrangement of
 \eqref{eq:collapse} permits the direct calculation of the free energy, assuming
 that the underlying physics of the regulatory architecture has not changed. Thus,
 it becomes possible to experimentally test the general assertions made in Fig.
 \ref{fig:deltaF_theory}.

\subsection*{DNA Binding Domain Mutations}
With this arsenal of analytic diagnostics, we can begin to explore the
mutational space of the repressor and map these mutations to the biophysical
parameters they control. As one of the most thoroughly studied transcription
factors, LacI has been subjected to numerous crystallographic and mutational
studies \cite{Daber2007, Daber2009a, Lewis1996, Swerdlow2014}. One such work
generated a set of point mutations in the LacI repressor and
examined the diversity of the phenotypic response to different allosteric
effectors \cite{Daber2011a}. However, experimental variables such as the
repressor copy number or the number of specific binding sites were not known,
making precise calculation of $\Delta F$ as presented here not tractable.
Using this dataset as a guide, we chose a subset of the mutations and
inserted them into our experimental strains of \textit{E. coli} where these parameters are known
and tightly controlled \cite{Garcia2011, Razo-Mejia2018}.

We made three amino acid substitutions (Y20I, Q21A, and Q21M) that are
critical for the DNA-repressor interaction. These mutations were introduced
into the \textit{lacI} sequence used in Garcia and Phillips 2011
\cite{Garcia2011} with four different ribosomal binding site sequences that
were shown (via quantitative Western blotting)
 to tune the wild-type
repressor copy number across three orders of magnitude. These mutant
constructs were integrated into the \textit{E. coli} chromosome harboring a
Yellow Fluorescent Protein (YFP) reporter. The YFP promoter included
the native O2 LacI operator sequence which the wild-type LacI repressor binds
with high specificity ($\Delta\varepsilon_{RA} = -13.9\, k_BT$). The
fold-change in gene expression for each mutant across twelve concentrations
of IPTG was measured via flow cytometry. As we mutated only a single amino
acid with the minimum number of base pair changes to the codons from the
wild-type sequence, we find it unlikely that the repressor copy number was
drastically altered from those reported in \cite{Garcia2011} for the
wild-type sequence paired with the same ribosomal binding site sequences. In
characterizing the effects of these DNA binding mutations, we take the
repressor copy number to be unchanged. Any error introduced by this mutation
should be manifest as a larger than predicted systematic shift in the free
energy change when the repressor copy number is varied.

\begin{SCfigure*}
\centering
\includegraphics{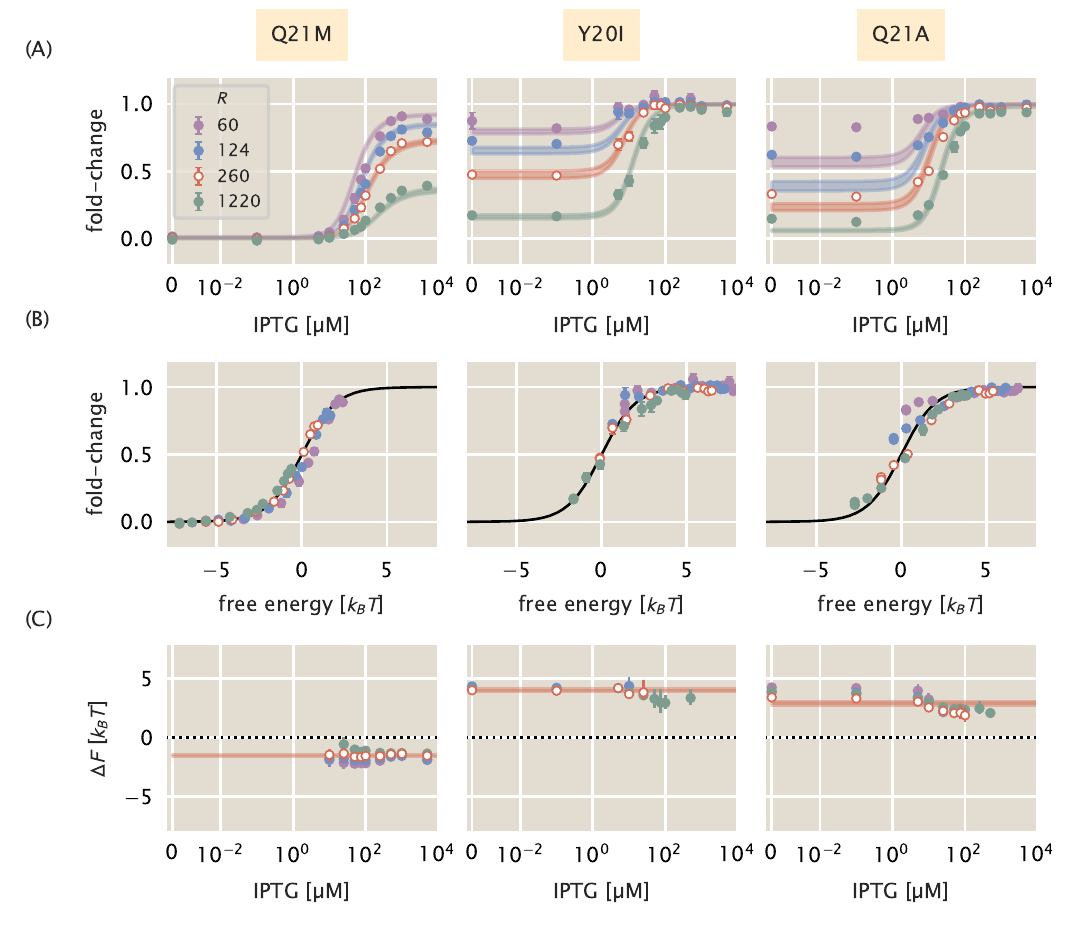}
\caption{Induction profiles and free energy modifications of DNA binding
domain mutations. Each column corresponds to the highlighted mutant at the top of the
figure. Each strain was paired with the native O2 operator sequence. White-faced points correspond to the strain for each mutant
from which the DNA binding energy was estimated.  (A) Induction profiles of
each mutant at four different repressor copy numbers as a function of the
inducer concentration. Points correspond to the mean fold-change in gene
expression of six to ten biological replicates. Error bars are the standard
error of the mean. Shaded regions demarcate the 95\% credible region of the
induction profile generated by the estimated DNA binding energy. (B) Data
collapse of all points for each mutant shown in (A) using only the DNA
binding energy estimated from a single repressor copy number. Points
correspond to the average fold-change in gene expression of six to ten
biological replicates. Error bars are standard error of the mean. Where error
bars are not visible, the relative error in measurement is smaller than the
size of the marker. (C) The change in the free energy resulting from each
mutation as a function of the inducer concentration. Points correspond to the
median of the marginal posterior distribution for the free energy. Error bars
represent the upper and lower bounds of the 95\% credible region. Points in
(A) at the detection limits of the flow cytometer (near fold-change values of
0 and 1) were neglected for calculation of the $\Delta F$. The IPTG
concentration is shown on a symmetric log scale with linear scaling ranging from
$0$ to $10^{-2}\,\mu$M and log scaling elsewhere.}
\label{fig:DNA_muts}
\end{SCfigure*}

A na\"{i}ve hypothesis for the effect of a mutation in the DNA binding domain is that
\textit{only} the DNA binding energy is altered. This hypothesis appears to
contradict the core principle of allostery in that ligand binding in one domain
influences binding in another, suggesting that changing
\text{any} parameter modifies them all. The characteristic curves summarized in Fig.
\ref{fig:deltaF_theory} give a means to  discriminate between these two
hypotheses by examining the change in the free energy. Using a single induction profile
(white-faced points in Fig. \ref{fig:DNA_muts}), we estimated the DNA binding
energy using a Bayesian approach, the details of which are discussed in the Materials and Methods as
well as the SI text. The shaded red region for each mutant in Fig.
\ref{fig:DNA_muts} represents the 95\% credible region of this fit whereas
all other shaded regions are 95\% credible regions of the predictions for other repressor copy
numbers. We find that redetermining only the DNA binding energy accurately
captures the majority of the induction profiles, indicating that other parameters
are unaffected. One exception is for the lowest
repressor copy numbers ($R = 60$ and $R=124$ per cell) of mutant Q21A
at low concentrations of IPTG. However, we note that this disagreement is
comparable to that observed for the wild-type repressor binding to the weakest
operator in Razo-Mejia \textit{et al.} 2018 \cite{Razo-Mejia2018}, illustrating that
our model is imperfect in characterizing weakly repressing architectures.
Including other parameters in the fit (such as $\Delta\varepsilon_{AI}$) does
not significantly improve the accuracy of the predictions. Furthermore, the magnitude of
this disagreement also depends on the choice of the fitting strain (see SI
text).

Mutations Y20I and Q21A both weaken the affinity of the repressor to the DNA 
relative to the wild type strain ($-9.9 ^{+0.1}_{-0.1}\, k_BT$ and
$-11.0^{+0.1}_{-0.1}\, k_BT$, respectively). Here we report the median of the
inferred posterior probability distribution with the superscripts and subscripts
corresponding to the upper and lower bounds of the 95\% credible region. 
These binding energies are comparable to that of the wild-type
repressor affinity to the native LacI operator sequence O3, with a DNA binding
energy of $-9.7\, k_BT$. The mutation Q21M increases the strength
of the DNA-repressor interaction relative to the wild-type repressor with a
binding energy of $-15.43^{+0.07}_{-0.06}\, k_BT$, comparable to the
affinity of the wild-type repressor to the native O1 operator sequence
($-15.3 k_BT$). It is notable that a single amino acid substitution
of the repressor is capable of changing the strength of the DNA binding
interaction well beyond that of many single base-pair mutations in the operator
sequence \cite{Barnes2018, Garcia2011}.

Using the new DNA binding energies, we can collapse all measurements of
fold-change as a function of the free energy as shown in Fig.
\ref{fig:DNA_muts}(B). This allows us to test the diagnostic power of the
decomposition of the free energy described in Fig. \ref{fig:deltaF_theory}. To 
compute the $\Delta F$ for each mutation, we inferred the observed
mean free energy of the mutant
strain for each inducer concentration and repressor copy number (see Materials and Methods
as well as the SI text for a detailed explanation of the inference). We note
that in the limit of extremely low or high fold-change, the inference of the
free energy is either over- or under-estimated, respectively, introducing a
systematic error. Thus, points which are close to these limits are omitted in
the calculation of $\Delta F$. We direct the reader to the SI text for a
detailed discussion of this systematic error. With a measure of $F\perst$
for each mutant at each repressor copy number, we compute the difference in free
energy relative to the wild-type strain with the same repressor copy number and operator sequence,
restricting all variability in $\Delta F$ solely to changes in
$\Delta\varepsilon_{RA}$. 

The change in free energy for each mutant is shown in Fig.
\ref{fig:DNA_muts}(C). It can be seen that the $\Delta F$ for each mutant is
constant as a function of the inducer concentration and is concordant with the
prediction generated from fitting $\Delta\varepsilon_{RA}$ to a single repressor
copy number [red lines Fig. \ref{fig:DNA_muts}(C)]. This is in line with the
predictions outlined in Fig. \ref{fig:deltaF_theory}(C) and (D), indicating that
the allosteric parameters are "insulated", meaning they are not affected by the
DNA binding domain mutations. As the $\Delta F$ for all repressor copy numbers collapses onto the
prediction, we can say that the expression of the repressor itself is the same
or comparable with that of the wild type. If the repressor copy number were perturbed in addition to $\Delta
\varepsilon_{RA}$, one would expect a shift away from the prediction that scales logarithmically
with the change in repressor copy number. However, as the $\Delta F$
is approximately the same for each repressor copy number, it can be surmised
that the mutation does not significantly change the expression or folding
efficiency of the repressor itself.  These results allow us to state that the DNA binding energy
$\Delta\varepsilon_{RA}$ is the only parameter modified by the DNA mutants
examined.

\subsection*{Inducer Binding Domain Mutations}
Much as in the case of the DNA binding mutants, we cannot safely assume
\textit{a priori} that
a given mutation in the inducer binding domain affects only the inducer
binding constants $K_A$ and $K_I$. While it is easy to associate the inducer
binding constants with the inducer binding domain, the critical parameter in
our allosteric model $\Delta\varepsilon_{AI}$ is harder to restrict to a
single spatial region of the protein. As $K_A$, $K_I$, and
$\Delta\varepsilon_{AI}$ are all parameters dictating the allosteric
response, we consider two hypotheses in which inducer binding mutations alter
either all three parameters or only $K_A$ and $K_I$.

We made four point mutations within the inducer binding domain of LacI (F164T, Q294V,
Q294R, and Q294K) that have been shown previously to alter binding to
multiple allosteric effectors \cite{Daber2011a}. In contrast to the DNA binding
domain mutants, we paired the inducer binding domain mutations with the three
native LacI operator sequences (which have various affinities for the repressor)
and a single ribosomal binding site sequence. This ribosomal binding site sequence, as reported in 
\cite{Garcia2011}, expresses the wild-type LacI repressor
to an average copy number of approximately $260$ per cell. As the free energy
differences resulting from point mutations in the DNA binding domain can be
described solely by changes to $\Delta\varepsilon_{RA}$, we continue under
the assumption that the inducer binding domain mutations do not significantly alter the repressor
copy number. 

\begin{figure*}[t]
        \centering
        \includegraphics{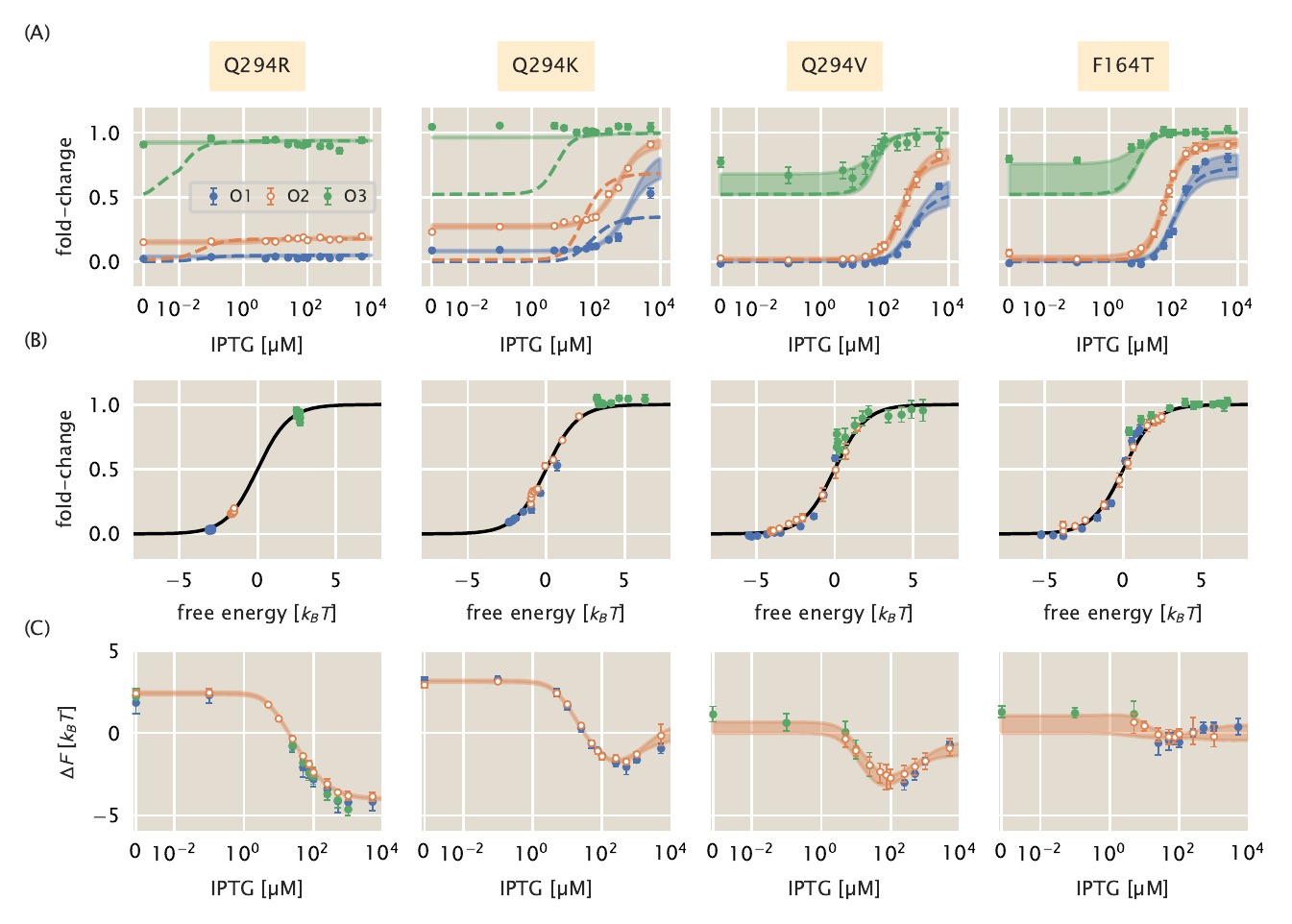}
        \caption{Induction profiles and free energy differences of inducer
        binding domain mutants. White faced points represent the strain to
        which the parameters were fit, namely the O2 operator sequence. Each column corresponds to the mutant
        highlighted at the top of the figure. All strains have $R = 260$ per
        cell. (A) The fold-change in gene expression as a function of the
        inducer concentration for three operator sequences of varying 
        strength. Dashed lines correspond to the curve of best fit resulting
        from fitting $K_A$ and $K_I$ alone. Shaded curves correspond to the
        95\% credible region of the induction profile determined from
        fitting $K_A$, $K_I$, and $\Delta\varepsilon_{AI}$. Points
        correspond to the mean measurement of six to twelve biological
        replicates. Error bars are the standard error of the mean. (B) Points
        in (A) collapsed as a function of the free energy calculated from
        redetermining $K_A$, $K_I$, and $\Delta\varepsilon_{AI}$. (C) Change
        in free energy resulting from each mutation as a function of the
        inducer concentration. Points correspond to the median of the
        posterior distribution for the free energy. Error bars represent the
        upper and lower bounds of the 95\% credible region. Shaded curves are
        the predictions. IPTG concentration is shown on a symmetric log scaling
        axis with the linear region spanning from $0$ to $10^{-2}\,\mu$M and log
        scaling elsewhere.}
        \label{fig:IND_muts}
\end{figure*}

The induction profiles for these four mutants are shown in Fig.
\ref{fig:IND_muts}(A). Of the mutations chosen, Q294R and Q294K appear to
have the most significant impact,  with Q294R abolishing the characteristic
sigmoidal titration curve entirely. It is notable that both Q294R and Q294K
have elevated expression in the absence of inducer compared to the other two
mutants paired with the same operator sequence. Panel (A) in Fig.
\ref{fig:deltaF_theory} illustrates that if only $K_A$ and $K_I$ were being
affected by the mutations, the fold-change should be identical for all mutants
in the absence of inducer. This discrepancy in the observed leakiness
immediately suggests that more than $K_A$ and $K_I$ are affected for Q294K
and Q294R.

Using a single induction profile for each mutant (shown in Fig.
\ref{fig:IND_muts} as white-faced circles), we inferred the parameter
combinations for both hypotheses and drew predictions for the induction
profiles with other operator sequences. We find that the simplest hypothesis (in
which only $K_A$ and $K_I$ are
altered) does not permit accurate prediction of most induction profiles.
These curves, shown as dotted lines in Fig. \ref{fig:IND_muts}(A), fail
spectacularly in
the case of Q294R and Q294K, and undershoot the observed profiles for F164T
and Q294V, especially when paired with the weak operator sequence O3. The
change in the leakiness for Q294R and Q294K is particularly evident as the
expression at $c = 0$ should be identical to the wild-type repressor under this hypothesis.
Altering only $K_A$ and $K_I$ is not sufficient to accurately predict the
induction profiles for F164T and Q294V, but not to the same degree as Q294K
and Q294R. The disagreement is most evident for the weakest operator O3
[green lines in Fig. \ref{fig:IND_muts}(A)], though we have discussed
previously that the induction profiles for weak operators are difficult to
accurately describe and can result in comparable disagreement for the
wild-type repressor \cite{Razo-Mejia2018, Barnes2018}.

Including $\Delta\varepsilon_{AI}$ as a perturbed parameter in addition to
$K_A$ and $K_I$ improves the predicted profiles for all four mutants. By
fitting these three parameters to a single strain, we are able to accurately
predict the induction profiles of other operators as seen by the shaded lines
in Fig. \ref{fig:IND_muts}(A). With these modified parameters, all
experimental measurements collapse as a function of their free energy as
prescribed by \eqref{eq:collapse} [Fig. \ref{fig:IND_muts}(B)]. All four
mutations significantly diminish the binding affinity of both states of the
repressor to the inducer, as seen by the estimated parameter values reported in
Tab. \ref{tab:ind_params}. As evident in the data alone, Q294R abrogates
inducibility outright ($K_A \approx K_I$).
For Q294K, the active state of the
repressor can no longer bind inducer whereas the inactive state binds with
weak affinity. The remaining two mutants, Q294V and F164T, both show
diminished binding affinity of the inducer to both the active and inactive
states of the repressor relative to the wild-type. 

\begin{table}[t]
\centering
    \caption{Inferred values of $K_A$, $K_I$, and $\Delta\varepsilon_{AI}$ for 
             inducer binding mutants}
    \begin{tabular}{lcccr}
    Mutant & $K_A$  & $K_I$  & $\Delta\varepsilon_{AI}$ [$k_BT$] & Reference \\
    \midrule
    WT & $139^{+29}_{-22}\,\mu$M & $0.53^{+0.04}_{-0.04}\,\mu$M & 4.5 & \cite{Razo-Mejia2018}\\
    &&&&\\
    F164T & $165^{+90}_{-65}\,\mu$M & $3^{+6}_{-3}\,\mu$M & $1^{+5}_{-2}$
    & This study\\
    &&&&\\
    Q294V & $650^{+450}_{-250}\,\mu$M & $8^{+8}_{-8}\,\mu$M &
    $3^{+6}_{-3}$ & This study\\
    &&&&\\
    Q294K & $> 1$ mM & $310^{+70}_{-60}\,\mu$M & $-3.11^{+0.07}_{-0.07}$ &
    This study\\
    &&&&\\
    Q294R & $9_{-9}^{+20}\,\mu$M & $8^{+20}_{-8}\,\mu$M & $-2.35^{+0.01}_{-0.09}$ & This study\\
    \bottomrule
    \label{tab:ind_params}
    \end{tabular}
\end{table} 

Given the collection of fold-change measurements, we computed the $\Delta F$
relative to the wild-type strain with the same operator and repressor copy
number. This leaves differences in $p_{act}(c)$ as the sole contributor to
the free energy difference, assuming our hypothesis that $K_A$, $K_I$, and
$\Delta\varepsilon_{AI}$ are the only perturbed parameters is correct. The
change in free energy can be seen in Fig. \ref{fig:IND_muts}(C). For all
mutants, the free energy difference inferred from the observed fold-change
measurements falls within error of the predictions generated under the
hypothesis that $K_A$, $K_I$, and $\Delta\varepsilon_{AI}$ are all affected
by the mutation [shaded curves in Fig. \ref{fig:IND_muts}(C)]. The profile of
the free energy change exhibits some of the rich phenomenology illustrated in
Fig. \ref{fig:deltaF_theory}(A) and (B). Q294K, F164T, and Q294V exhibit a
non-monotonic dependence on the inducer concentration, a feature that can
only appear when $K_A$ and $K_I$ are altered. The non-zero $\Delta F$ at
$c=0$ for Q294R and Q294K coupled with an inducer concentration dependence is
a telling sign that $\Delta\varepsilon_{AI}$ must be significantly modified.
This shift in $\Delta F$ is positive in all cases, indicating that
$\Delta\varepsilon_{AI}$ must have decreased, and that the inactive state
has become more energetically favorable for these mutants than for the wild-type protein. Indeed
the estimates for $\Delta\varepsilon_{AI}$ (Tab. \ref{tab:ind_params})
reveal both mutations Q294R and Q294K make
the inactive state more favorable than the active state. Thus, for these two
mutations, only $\approx 10\%$ of the repressors are active in the absence of
inducer, whereas the basal active fraction is $\approx 99\%$ for the wild-type
repressor \cite{Razo-Mejia2018}.

Taken together, these parametric changes diminish the response of the regulatory
architecture as a whole to changing inducer concentrations. They furthermore
reveal that the parameters which govern the allosteric response are 
interdependent and no single parameter is insulated from the others. However, as
\textit{only} the allosteric parameters are changed, one can
say that the allosteric parameters as a whole are insulated from the other
components which define the regulatory response, such as repressor copy number
and DNA binding affinity.

\subsection*{Predicting Effects of Pairwise Double Mutations}
Given full knowledge of each individual mutation, we can draw predictions of the
behavior of the pairwise double mutants with no free parameters based on the
simplest null hypothesis of no epistasis. The formalism of $\Delta F$ defined by
\eqref{eq:delF} explicitly states that the contribution to the free energy
of the system from the difference in DNA binding energy and the allosteric parameters are
strictly additive. Thus, deviations from the predicted change in free energy
would suggest epistatic interactions between the two mutations.

\begin{figure*}
    \centering
    \includegraphics{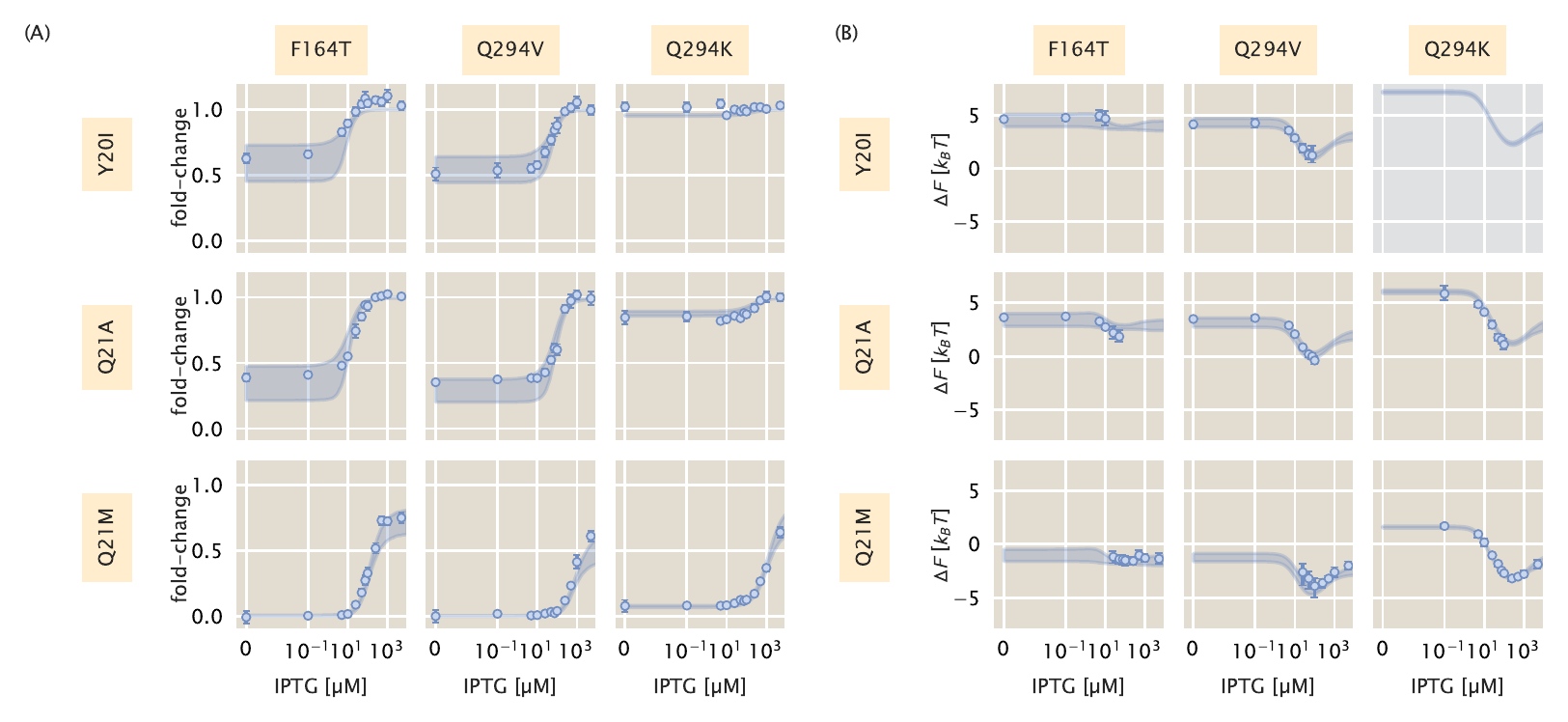}
    \caption{Induction and free energy profiles of DNA binding and 
        inducer binding double mutants.(A) Fold-change in gene expression for
        each double mutant as a function of IPTG. Points and errors correspond
        to the mean and standard error of six to ten biological replicates.
        Where not visible, error bars are smaller than the corresponding marker. Shaded
        regions correspond to the 95\% credible region of the prediction given
        knowledge of the single mutants. These were generated by drawing 10$^4$
        samples from the $\Delta\varepsilon_{RA}$ posterior distribution of the
        single DNA binding domain mutants and the
        joint probability distribution of $K_A$, $K_I$, and
        $\Delta\varepsilon_{AI}$ from the single inducer binding domain mutants.
        (B) The difference in free energy of each double mutant as a function of
        the reference free energy. Points and errors correspond to the median
        and bounds of the 95\% credible region of the posterior distribution for
        the inferred $\Delta F$. Shaded lines region are the predicted change in
        free energy, generated in the same manner as the shaded lines in (A).
        All measurements were taken from a strain with 260 repressors per cell
        paired with a reporter with the native O2 LacI operator sequence. In
        all plots, the IPTG concentration is shown on a symmetric log axis with
        linear scaling between $0$ and $10^{-2}\,\mu$M and log scaling elsewhere.}
    \label{fig:dbl_muts}
\end{figure*}

To test this additive model, we constructed nine double mutant strains, each
having a unique inducer binding (F164T, Q294V, Q294K) and DNA binding
mutation (Y20I, Q21A, Q21M). To make predictions with an appropriate
representation of the uncertainty, we computed a large array of induction
profiles given random draws from the posterior distribution for the DNA binding
energy (determined from the single DNA binding mutants) as well as from the
joint posterior for the allosteric parameters (determined from the single
inducer binding mutants). These predictions, shown in Fig.
\ref{fig:dbl_muts}(A) and (B) as shaded blue curves, capture all
experimental measurements of the fold-change [Fig. \ref{fig:dbl_muts}(A)] and
the inferred difference in free energy [Fig. \ref{fig:dbl_muts}(B)]. The
latter indicates that there are no epistatic interactions between the
mutations queried in this work, though if there were, systematic deviations from these
predictions would shed light on how the epistasis is manifest. 

The precise agreement between the predictions and measurements for Q294K
paired with either Q21A or Q21M is striking as Q294K drastically changed
$\Delta\varepsilon_{AI}$ in addition to $K_A$ and $K_I$. Our ability to
predict the induction profile and free energy change underscores the extent
to which the DNA binding energy and the allosteric parameters are insulated
from one another. Despite this insulation, the repressor still functions as
an allosteric molecule, emphasizing that the mutations we have inserted do not
alter the pathway of communication between the two domains of
the protein. 
As the double mutant Y20I-Q294K exhibits fold-change of approximately $1$
across all IPTG concentrations [Fig. \ref{fig:dbl_muts}(A)], these mutations
in tandem make repression so weak it is beyond the limits which are
detectable by our experiments. As a consequence, we are unable to estimate
$\Delta F$ nor experimentally verify the corresponding prediction [grey box
in Fig. \ref{fig:dbl_muts}(B)]. However, as the predicted fold-change in gene
expression is also approximately $1$ for all $c$, we believe that the
prediction shown for $\Delta F$ is likely accurate. One would be able to
infer the $\Delta F$ to confirm these predictions using a more sensitive
method for measuring the fold-change, such as single-cell microscopy or
colorimetric assays.

\subsection*{Discussion}
Allosteric regulation is often couched as ``biological action at a distance".
Despite extensive knowledge of protein structure and function, it remains
difficult to translate the coordinates of the atomic constituents of a
protein to the precise parameter values which define the functional response,
making each mutant its own intellectual adventure.
Bioinformatic approaches to understanding the sequence-structure relationship 
have permitted us to examine how the residues of allosteric proteins
evolve, revealing conserved regions which hint to their function.
Co-evolving residues reveal sectors of conserved
interactions which traverse the protein that act as the allosteric
communication channel between domains \cite{Suel2002, McLaughlin2012,
Reynolds2011}. Elucidating these sectors has advanced our understanding of
how distinct domains "talk" to one another and has permitted direct
engineering of allosteric responses into non-allosteric enzymes
\cite{Raman2014, Raman2016, Poelwijk2016}. Even so, we are left without a
quantitative understanding of how these admittedly complex networks set the
energetic difference between active and inactive states or how a given mutation
influences binding affinity. In this context, a biophysical model
in which the various parameters are intimately connected to the molecular
details can be of use and can lead to quantitative predictions of
the interplay between amino-acid identity and system-level response.

By considering how each parameter contributes to the observed change in free
energy, we are able to tease out different classes of parameter perturbations
which result in stereotyped responses to changing inducer concentration.
These characteristic changes to the free energy can be used as a diagnostic
tool to classify mutational effects. For example, we show in Fig.
\ref{fig:deltaF_theory} that modulating the inducer binding constants $K_A$
and $K_I$ results in non-monotonic free energy changes that are dependent on
the inducer concentration, a feature observed in the inducer binding mutants
examined in this work. Simply looking at the inferred $\Delta F$ as a function
of inducer concentration, which
requires no fitting of the biophysical parameters, indicates that $K_A$ and $K_I$ must be
modified considering those are the only parameters which can generate such a response.

Another key observation is that a perturbation to only $K_A$ and $K_I$ requires
that the $\Delta F = 0$ at $c = 0$. Deviations from this condition imply that more than
the inducer binding constants must have changed. If this shift in $\Delta F$ off of $0$ at
$c = 0$ is not constant across all inducer concentrations, we can surmise
that the energy difference between the allosteric states
$\Delta\varepsilon_{AI}$ must also be modified. We again see this effect for all
of our inducer mutants. By examining the inferred $\Delta F$, we can
immediately say that in addition to $K_A$ and $K_I$, $\Delta\varepsilon_{AI}$
must decrease relative to the wild-type value as $\Delta F > 0$ at $c = 0$.
When the allosteric parameters are fit to the induction profiles, we indeed
see that this is the case, with all four mutations decreasing the energy gap
between the active and inactive states. Two of these mutations, Q294R and
Q294K, make the inactive state of the repressor \textit{more} stable than the
active state, which is not the case for the wild-type
repressor \cite{Razo-Mejia2018}.

Our formulation of $\Delta F$ indicates that shifts away from $0$ that are
independent of the inducer concentration can only arise from changes to the
repressor copy number and/or DNA binding specificity, indicating that the
allosteric parameters are untouched. We see that for three mutations in the
DNA binding domain, $\Delta F$ is the same irrespective of the inducer
concentration. Measurements of $\Delta F$ for these mutants with repressor
copy numbers across three orders of magnitude yield approximately the same
value, revealing that $\Delta\varepsilon_{RA}$ is the sole parameter altered
via the mutations.

We note that the conclusions stated above can be qualitatively drawn without
resorting to fitting various parameters and measuring the goodness-of-fit.
Rather, the distinct behavior of $\Delta F$ is sufficient to determine which
parameters are changing. Here, these conclusions are quantitatively confirmed by
fitting these parameters to the induction profile, which results in accurate
predictions of the fold-change and $\Delta F$ for nearly every strain across 
different mutations, repressor copy numbers, and operator sequence, all at
different inducer concentrations. With a collection of evidence as to what
parameters are changing for single mutations, we put our model to the test
and drew predictions of how double mutants would behave both in terms of the
titration curve and free energy profile.

A hypothesis that arises from our formulation of $\Delta F$ is that a
simple summation of the energetic contribution of each mutation should be
sufficient to predict the double mutants (so long as they are in separate
domains). We find that such a calculation permits precise and accurate
predictions of the double mutant phenotypes, indicating that there are no
epistatic interactions between the mutations examined in this work. With an
expectation of what the free energy differences should be, epistatic
interactions could be understood by looking at how the measurements deviate
from the prediction. For example, if epistatic interactions exist which
appear as a systematic shift from the predicted $\Delta F$ 
independent of inducer concentration, one could conclude that DNA binding
energy is not equal to that of the single mutation in the DNA binding domain alone. Similarly,
systematic shifts that are dependent on the inducer concentration (i.e. not
constant) indicate that the allosteric parameters must be influenced. If the
expected difference in free energy is equal to $0$ when $c=0$, one could
surmise that the modified parameter must not be $\Delta\varepsilon_{AI}$ nor
$\Delta\varepsilon_{RA}$ as these would both result in a shift in leakiness,
indicating that $K_A$ and $K_I$ are further modified.

\begin{figure}
    \centering
    \includegraphics{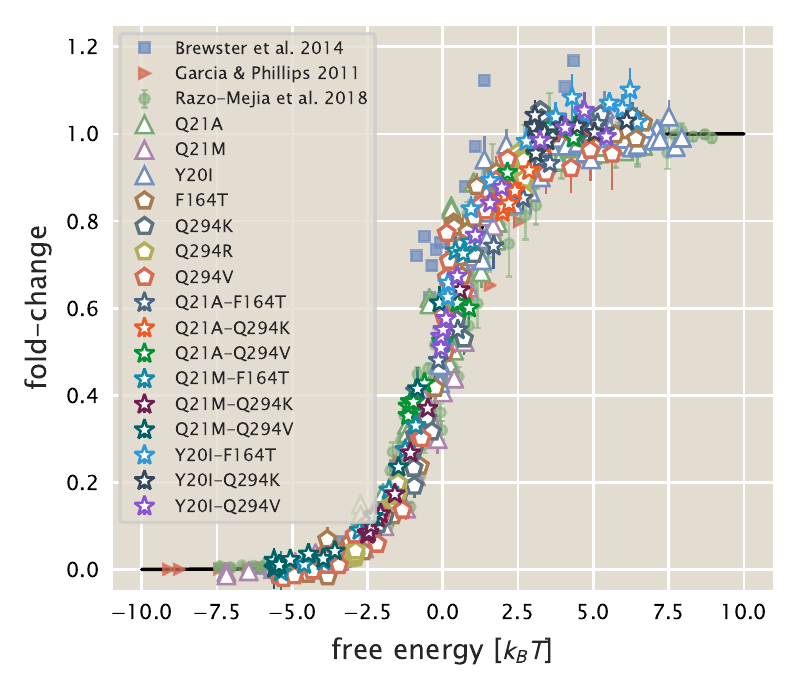}
    \caption{Data collapse of the simple repression regulatory architecture.
    All data are means of biological replicates. Where present, error bars
    correspond to the standard error of the mean of five to fifteen
    biological replicates. Red triangles indicate data from Garcia and
    Phillips \cite{Garcia2011} obtained by colorimetric assays. Blue squares
    are data from Brewster et al.\cite{Brewster2014} acquired from video
    microscopy. Green circles are data from Razo-Mejia et al.
    \cite{Razo-Mejia2018} obtained via flow cytometry. All other symbols
    correspond to the work presented here. An interactive version of this
    figure can be found on the
    \href{https://www.rpgroup.caltech.edu/mwc_mutants}{paper website} where the different data sets can be viewed in more detail.}
    \label{fig:all_data_collapse}
\end{figure}

Ultimately, we present this work as a proof-of-principle for using
biophysical models to investigate how mutations influence the response of
allosteric systems. We emphasize that such a treatment allows one to boil
down the complex phenotypic responses of these systems to a single-parameter
description which is easily interpretable as a free energy. The general
utility of this approach is illustrated in Fig. \ref{fig:all_data_collapse}
where gene expression data from previous work \cite{Garcia2011, Brewster2014,
Razo-Mejia2018} along with all of the measurements presented in this work
collapse onto the master curve defined by \eqref{eq:collapse}. While our
model coarse grains many of the intricate details of transcriptional
regulation into two states (one in which the repressor is bound to the
promoter and one where it is not), it is sufficient to describe a wide range
of regulatory scenarios. Given enough parametric knowledge of the system, it
becomes possible to examine how modifications to the parameters move the
physiological response along this reduced one-dimensional parameter space.
This approach offers a glimpse at how mutational effects can be described in terms of
energy rather than Hill coefficients and arbitrary prefactors. While we have
explored a very small region of sequence space in this work, coupling of this
approach with high-throughput sequencing-based methods to query a library of
mutations within the protein will shed light on the phenotypic landscape
centered at the wild-type sequence. Furthermore, pairing libraries of protein
and operator sequence mutants will provide insight as to how the protein and
regulatory sequence coevolve, a topic rich with opportunity for a dialogue
between theory and experiment.

\matmethods{
    \subsection*{Bacterial Strains and DNA Constructs}
    All wild-type strains from which the mutants were derived were generated in
    previous work from the Phillips group \cite{Garcia2011, Razo-Mejia2018}.
    Briefly, mutations were first introduced into the {\it lacI} gene of our
    pZS3*1-lacI plasmid \cite{Garcia2011} using a combination of overhang PCR
    Gibson assembly as well as QuickChange mutagenesis (Agligent Technologies).
    The oligonucleotide sequences used to generate each mutant as well as the
    method are provided in the SI text.
    
    For mutants generated through overhang PCR and Gibson assembly,
    oligonucleotide primers were purchased containing an overhang with the
    desired mutation and used to amplify the entire plasmid. Using the homology
    of the primer overhang, Gibson assembly was performed to circularize the DNA
    prior to electroporation into MG1655 {\it E. coli} cells. Integration of LacI
    mutants was performed with $\lambda$ Red recombineering \cite{Sharan2009}
    as described in Reference \cite{Garcia2011}.
    
    The mutants studied in this work were chosen from data reported in
    \cite{Daber2011a}. In selecting mutations, we looked for mutants which suggested
    moderate to strong deviations from the behavior of the wild-type repressor. We
    note that the variant of LacI used in this work has an additional three amino
    acids (Met-Val-Asn) added to the N-terminus than the canonical LacI sequence
    reported in \cite{Farabaugh1978}. For this reason, all mutants given here are
    with respect to our sequence and their positions are shifted by three to those
    studied in \cite{Daber2011a}.
    
    \subsection*{Flow Cytometry}
    All fold-change measurements were performed on a MACSQuant flow cytometer as
    described in Razo-Mejia et al. \cite{Razo-Mejia2018}. Briefly, saturated
    overnight cultures 500 $\mu$L in volume were grown in deep-well 96 well plates 
    covered with  a breathable nylon cover (Lab Pak - Nitex
    Nylon, Sefar America, Cat. No. 241205). After approximately 12 to 15 hr, the
    cultures reached saturation and were diluted 1000-fold into a second 2 mL
    96-deep-well plate where each well contained 500 $\mu$L of M9 minimal media
    supplemented with 0.5\% w/v glucose (anhydrous D-Glucose, Macron Chemicals)
    and the appropriate concentration of IPTG (Isopropyl
    $\beta$-D-1-thiogalactopyranoside, Dioxane Free, Research Products International).
    These were sealed with a breathable cover and were allowed to grow for
    approximately 8 hours until the OD$_\text{600nm} \approx 0.3$. 
    Cells were then diluted ten-fold into a round-bottom 96-well plate 
    (Corning Cat. No. 3365) containing 90 $\mu$L of M9 minimal media supplemented 
    with 0.5\% w/v glucose along with the corresponding IPTG concentrations.
    
    The flow cytometer was calibrated prior to use with MACSQuant Calibration
    Beads (Cat. No. 130-093-607). During measurement, the cultures were held at
    approximately 4$^\circ$ C by placing the 96-well plate on a MACSQuant ice block.
    All fluorescence measurements were made using a 488 nm excitation wavelength
    with a 525/50 nm emission filter. The photomultiplier tube voltage settings for
    the instrument are the same as those used in Reference \cite{Razo-Mejia2018}.

    The data was processed using an automatic unsupervised gating procedure based
    on the front and side- scattering values, where we fit a two-dimensional
    Gaussian function to the $\log_{10}$ forward-scattering (FSC) and the
    $\log_{10}$ side-scattering (SSC) data. Here we assume that the region with
    highest density of points in these two channels corresponds to single-cell
    measurements and consider data points that fall within 40\% of the highest
    density region of the two-dimensional Gaussian function. We direct the reader
    to Reference \cite{Razo-Mejia2018} for further detail and comparison of flow
    cytometry with single-cell microscopy.
    
    \subsection*{Bayesian Parameter Estimation}
    We used a Bayesian definition of probability in the statistical analysis of
    all mutants in this work. In the SI text, we derive
    in detail the statistical models used for the various parameters as well as
    multiple diagnostic tests. Here, we give a generic description of our
    approach. To be succinct in notation, we consider a generic parameter
    $\theta$ which represents $\Delta\varepsilon_{RA}$, $K_A$, $K_I$, and/or
    $\Delta\varepsilon_{AI}$ depending on the specific LacI mutant.
    
    As prescribed by
    Bayes' theorem, we are interested in the posterior probability distribution
    \begin{equation}
        g(\theta\,\vert\, y) \propto {f(y\,\vert\,\theta)g(\theta)},
        \label{eq:bayes_generic}
    \end{equation}
    where we use $g$ and $f$ to represent probability densities over parameters and
    data, respectively, and $y$ to represent a set of fold-change measurements. The
    likelihood of observing our dataset $y$ given a value of $\theta$ is captured by
    $f(y\,\vert\,\theta)$. All prior information we have about the possible values
    of $\theta$ are described by $g(\theta)$. 
    
    In all inferential models used in this work, we assumed that all experimental measurements at a
    given inducer concentration were normally distributed about a mean value $\mu$
    dictated by \eqref{eq:foldchange} with a variance $\sigma^2$, 
    \begin{equation}
        f(y\,\vert\, \theta) = {1 \over (2\pi\sigma^2)^{N/2}}\prod\limits_i^N \exp\left[-{(y_i - \mu(\theta))^2 \over 2\sigma^2}\right],
        \label{eq:generic_likelihood}
    \end{equation}
    where $N$ is the number of measurements in the data set $y$.
    
    This choice of likelihood is justified as each individual measurement at a given
    inducer concentration is a biological replicate and independent of all other
    experiments. By using a Gaussian likelihood, we introduce another parameter
    $\sigma$. As $\sigma$ must be positive and greater than zero, we define as a prior
    distribution  a half-normal distribution with a standard deviation $\phi$, 
    \begin{equation}
        g(\sigma) = {{1 \over \phi}\sqrt{2 \over \pi}}\exp\left[-{x \over 2\phi^2}\right]\,;\, x \geq 0 
        \label{eq:sigma_prior},
    \end{equation}
    where $x$ is a given range of values for $\sigma$. A standard deviation of
    $\phi=0.1$ was chosen given our knowledge of the scale of our measurement
    error from other experiments. As the absolute measurement of fold-change is
    restricted between $0$ and $1.0$, and given our knowledge of the sensitivity
    of the experiment, it is reasonable to assume that the error will be closer
    to $0$ than to $1.0$. Further justification of this choice of prior through
    simulation based methods are given in the SI text. The prior distribution for
    $\theta$ is dependent on the parameter and its associated physical and
    physiological restrictions. Detailed discussion of our chosen prior
    distributions for each model can also be found in the SI text.
    
    All statistical modeling and parameter inference was performed using Markov
    chain Monte Carlo (MCMC). Specifically, Hamiltonian Monte Carlo sampling was
    used as sis implemented in the Stan probabilistic programming language \cite{Carpenter2017}. All
    statistical models saved as \texttt{.stan} models and can be accessed at the
    \href{https://www.github.com/rpgroup-pboc/mwc_mutants}{GitHub repository}
    associated with this work (DOI: 10.5281/zenodo.2721798) or can be downloaded directly from the
    \href{https://www.rpgroup.caltech.edu/mwc_mutants}{paper website}. 
    
    \subsection*{Inference of Free Energy From Fold-Change Data} While the
    fold-change in gene expression is restricted to be between 0 and 1, experimental
    noise can generate fold-change measurements beyond these bounds. To determine
    the free energy for a given set of fold-change measurements (for one unique
    strain at a single inducer concentration), we modeled the observed fold-change
    measurements as being drawn from a normal distribution with a mean $\mu$
    and standard deviation $\sigma$. Using Bayes' theorem, we can write
    the posterior distribution as 
    \begin{equation}
        g(\mu, \sigma\,\vert y) \propto
        g(\mu)g(\sigma){1\over(2\pi\sigma^2)^{N/2}}\prod\limits_i^N
        \exp\left[{-(y_i - \mu)^2 \over 2\sigma^2}\right]
        \label{eq:post_fc_mu},
    \end{equation}
    where $y$ is a collection of fold-change measurements. The prior distribution
    for $\mu$ was chosen to be uniform between 0 and 1 while the prior on $\sigma$
    was chosen to be half normal, as written in \eqref{eq:sigma_prior}. The
    posterior distribution was sampled independently for each set of fold-change
    measurements using MCMC. The \texttt{.stan} model for this inference is available on the
    \href{http://www.rpgroup.caltech.edu/mwc_mutants}{paper website}. 
    
    For each MCMC sample of $\mu$, the free energy was calculated as 
    \begin{equation}
        F = -\log\left(\mu^{-1} - 1\right)
        \label{eq:empirical_F}
    \end{equation}
    which is simply the rearrangement of \eqref{eq:collapse}. Using simulated
    data, we determined that when $\mu < \sigma$ or $(1 - \mu) < \sigma$, the
    mean fold-change in gene expression was over or underestimated for the lower
    and upper limit, respectively. This means that there are maximum and minimum
    levels of fold-change that can be detected using flow cytometry which are set
    by the distribution of fold-change measurements resulting from various
    sources of day-to-day variation. This results in a systematic error in the
    calculation of the free energy, making proper inference beyond these limits
    difficult. This bounds the range in which we can confidently infer this
    quantity with flow cytometry. We hypothesize that more sensitive methods,
    such as single cell microscopy, colorimetric assays, or direct counting of
    mRNA transcripts via Fluorescence \textit{In Situ} Hybridization (FISH) would
    improve the measurement of $\Delta F$. We further discuss details of this
    limitation in the SI text.
    
    \subsection*{Data and Code Availability}
    All data was collected, stored, and preserved using the Git version control
    software. Code for data processing, analysis, and figure generation is
    available on the GitHub repository
    (\href{https://www.github.com/rpgroup-pboc/mwc_mutants}{https://www.github.com/rpgroup-pboc/mwc\_mutants})
    or can be accessed via the
    \href{http://www.rpgroup.caltech.edu/mwc_mutants}{paper website}. Raw flow
    cytometry data is stored on the CaltechDATA data repository and can be
    accessed via DOI 10.22002/D1.1241.

}
\showmatmethods{} 

\acknow{
    We thank Pamela Bj\"{o}rkman, Rachel Galimidi, and Priyanthi Gnanapragasam
    for access and training for the use of the Miltenyi Biotec MACSQuant flow
    cytometer. The experimental efforts first took place at the Physiology summer
    course at the Marine Biological Laboratory in Woods Hole, MA, operated by the
    University of Chicago. We thank Ambika Nadkarni and Damian Dudka for their
    work on the project during the course. We also thank Suzannah Beeler, Justin
    Bois, Robert Brewster, Soichi Hirokawa, Heun Jin Lee, and Muir Morrison for
    thoughtful advice and discussion. This work was supported by La Fondation
    Pierre-Gilles de Gennes, the Rosen Center at Caltech, the NIH DP1 OD0002179
    (Director's Pioneer Award), R01 GM085286, and 1R35 GM118043 (MIRA). Nathan M.
    Belliveau was supported by a Howard Hughes Medical Institute International
    Student Research fellowship.    
}
\showacknow{} 

\bibliography{library}

\begin{thebibliography}{10}

\bibitem{Ackers1982}
Ackers GK, Johnson AD, Shea MA (1982) {Quantitative model for gene regulation
  by lambda phage repressor.}
\newblock {\em Proceedings of the National Academy of Sciences of the United
  States of America} 79(4):1129--33.

\bibitem{Buchler2003}
Buchler NE, Gerland U, Hwa T (2003) {On schemes of combinatorial transcription
  logic.}
\newblock {\em Proceedings of the National Academy of Sciences of the United
  States of America} 100(9):5136--41.

\bibitem{Vilar2003}
Vilar JM, Leibler S (2003) {DNA Looping and Physical Constraints on
  Transcription Regulation}.
\newblock {\em Journal of Molecular Biology} 331(5):981--989.

\bibitem{Garcia2011}
Garcia HG, Phillips R (2011) {Quantitative dissection of the simple repression
  input-output function.}
\newblock {\em Proceedings of the National Academy of Sciences of the United
  States of America} 108(29):12173--8.

\bibitem{Daber2011a}
Daber R, Sochor MA, Lewis M (2011) {Thermodynamic analysis of mutant lac
  repressors}.
\newblock {\em Journal of Molecular Biology} 409(1):76--87.

\bibitem{Brewster2014}
Brewster RC, et~al. (2014) {The transcription factor titration effect dictates
  level of gene expression}.
\newblock {\em Cell} 156(6):1312--1323.

\bibitem{Weinert2014}
Weinert FM, Brewster RC, Rydenfelt M, Phillips R, Kegel WK (2014) {Scaling of
  Gene Expression with Transcription-Factor Fugacity}.
\newblock {\em Physical Review Letters} 113(25):1--5.

\bibitem{Rydenfelt2014}
Rydenfelt M, Garcia HG, Cox RS, Phillips R (2014) {The influence of promoter
  architectures and regulatory motifs on gene expression in \textit{Escherichia
  coli}}.
\newblock {\em PLoS ONE} 9(12):1--31.

\bibitem{Razo-Mejia2014}
Razo-Mejia M, et~al. (2014) {Comparison of the theoretical and real-world
  evolutionary potential of a genetic circuit}.
\newblock {\em Physical biology} 11(2):026005.

\bibitem{Razo-Mejia2018}
Razo-Mejia M, et~al. (2018) {Tuning Transcriptional Regulation through
  Signaling: A Predictive Theory of Allosteric Induction}.
\newblock {\em Cell Systems} 6(4):456--469.

\bibitem{Bintu2005}
Bintu L, et~al. (2005) {Transcriptional regulation by the numbers: models}.
\newblock {\em Current Opinion in Genetics {\&} Development} 15(2):116--124.

\bibitem{Bintu2005a}
Bintu L, et~al. (2005) {Transcriptional regulation by the numbers:
  Applications}.
\newblock {\em Current Opinion in Genetics and Development} 15(2):125--135.

\bibitem{Kuhlman2007}
Kuhlman T, Zhang Z, Saier MH, Hwa T (2007) {Combinatorial transcriptional
  control of the lactose operon of \textit{Escherichia coli}.}
\newblock {\em Proceedings of the National Academy of Sciences of the United
  States of America} 104(14):6043--6048.

\bibitem{Swem2008}
Swem LR, Swem DL, Wingreen NS, Bassler BL (2008) {Deducing Receptor Signaling
  Parameters from In Vivo Analysis: LuxN/AI-1 Quorum Sensing in \textit{Vibrio
  harveyi}}.
\newblock {\em Cell} 134(3):461--473.

\bibitem{Keymer2006}
Keymer JE, Endres RG, Skoge M, Meir Y, Wingreen NS (2006) {Chemosensing in
  \textit{Escherichia coli}: Two regimes of two-state receptors.}
\newblock {\em Proceedings of the National Academy of Sciences of the United
  States of America} 103(6):1786--1791.

\bibitem{Phillips2015}
Phillips R, Milo R (2015) {Rates and duration} in {\em Cell Biology by the
  Numbers}.

\bibitem{Frumkin2018}
Frumkin I, et~al. (2018) Codon usage of highly expressed genes affects
  proteome-wide translation efficiency.
\newblock {\em Proceedings of the National Academy of Sciences}
  115(21):E4940--E4949.

\bibitem{Daber2007}
Daber R, Stayrook S, Rosenberg A, Lewis M (2007) {Structural Analysis of Lac
  Repressor Bound to Allosteric Effectors}.
\newblock {\em Journal of Molecular Biology} 370(4):609--619.

\bibitem{Daber2009a}
Daber R, Lewis M (2009) {Towards evolving a better repressor}.
\newblock {\em Protein Engineering, Design and Selection} 22(11):673--683.

\bibitem{Lewis1996}
Lewis M, et~al. (1996) {Crystal Structure of the Lactose Operon Repressor and
  Its Complexes with DNA and Inducer}.
\newblock {\em Science} 271(5253).

\bibitem{Swerdlow2014}
Swerdlow SJ, Schaaper RM (2014) {Mutagenesis in the lacI gene target of
  \textit{E. coli}: improved analysis for lacI(d) and lacO mutants.}
\newblock {\em Mutation research} 770:79--84.

\bibitem{Barnes2018}
Barnes SL, Belliveau NM, Ireland WT, Kinney JB, Phillips R (2019) {Mapping DNA
  sequence to transcription factor binding energy in vivo}.
\newblock {\em PLoS Comput Biol} 15(2).

\bibitem{Suel2002}
S{\"u}el GM, Lockless SW, Wall MA, Ranganathan R (2002) {Evolutionarily
  conserved networks of residues mediate allosteric communication in proteins}.
\newblock {\em Nature Structural Biology} 10(1):59--69.

\bibitem{McLaughlin2012}
{McLaughlin Jr} RN, Poelwijk FJ, Raman A, Gosal WS, Ranganathan R (2012) {The
  spatial architecture of protein function and adaptation}.
\newblock {\em Nature} 491(7422):138--142.

\bibitem{Reynolds2011}
Reynolds KA, McLaughlin RN, Ranganathan R (2011) {Hot Spots for Allosteric
  Regulation on Protein Surfaces}.
\newblock {\em Cell} 147(7):1564--1575.

\bibitem{Raman2014}
Raman S, Taylor N, Genuth N, Fields S, Church GM (2014) {Engineering
  allostery}.
\newblock {\em Trends in Genetics} 30(12):521--8.

\bibitem{Raman2016}
Raman AS, White KI, Ranganathan R (2016) {Origins of Allostery and Evolvability
  in Proteins: A Case Study.}
\newblock {\em Cell} 166(2):468--80.

\bibitem{Poelwijk2016}
Poelwijk FJ, Krishna V, Ranganathan R (2016) The {{Context}}-{{Dependence}} of
  {{Mutations}}: {{A Linkage}} of {{Formalisms}}.
\newblock {\em PLOS Computational Biology} 12(6):e1004771.

\bibitem{Sharan2009}
Sharan SK, Thomason LC, Kuznetsov SG, Court DL (2009) {Recombineering: a
  homologous recombination-based method of genetic engineering}.
\newblock {\em Nature Protocols} 4(2):206--223.

\bibitem{Farabaugh1978}
Farabaugh PJ (1978) Sequence of the {\textit{laci}} gene.
\newblock {\em Nature} 274(24):5.

\bibitem{Carpenter2017}
Carpenter B, et~al. (2017) Stan: A probabilistic programming language.
\newblock {\em Journal of Statistical Software, Articles} 76(1):1--32.

\end{thebibliography}

\end{document}